\documentclass[3p,times]{elsarticle}

\usepackage{ecrc}

\usepackage{setspace}

\volume{00}

\firstpage{1}

\journalname{Applied Mathematical Modelling}

\jnltitlelogo{AMM}

\usepackage{amssymb}
\usepackage{amsfonts}
\usepackage{amsmath}
\usepackage{graphicx}
\usepackage{epstopdf}
\usepackage{subcaption}
\usepackage{subfig}
\usepackage{ulem}
\usepackage{mathtools}
\usepackage[autostyle]{csquotes}
\usepackage{color}
\usepackage{amsthm}

\usepackage{lineno}

\usepackage[figuresright]{rotating}

\begin{document}

\begin{frontmatter}


\title{Dynamics of Spatial Heterogeneity in a Landfill with Interacting Phase Densities - A Stochastic Analysis}

\author{Amit K Chattopadhyay \footnote{Corresponding author}}
\ead{a.k.chattopadhyay@aston.ac.uk}
\ead[http://www.aston.ac.uk/eas/staff/a-z/amit-k-chattopadhyay/]{home page}
\address{Nonlinearity and Complexity Research Group, Aston University, Birmingham B4 7ET, England}
\author{Prasanta K Dey}
\address{Aston Business School, Aston University, Birmingham B4 7ET, England}
\author{Sadhan K Ghosh}
\address{Mechanical Engineering Department, Centre for Quality Management System, Jadavpur University, Kolkata 700 032, India}

\dochead{}





\begin{abstract}
A landfill represents a complex and dynamically evolving structure that can be stochastically perturbed by exogenous factors. Both thermodynamic (equilibrium) and time varying (non-steady state) properties of a landfill are affected by spatially heterogenous and nonlinear subprocesses that combine with constraining initial and boundary conditions arising from the associated surroundings. While multiple approaches have been made to model landfill statistics by incorporating spatially dependent parameters on the one hand (data based approach) and continuum dynamical mass-balance equations on the other (equation based modelling), practically no attempt has been made to amalgamate these two approaches while also incorporating inherent stochastically induced fluctuations affecting the process overall. In this article, we will implement a minimalist scheme of modelling the time evolution of a realistic three dimensional landfill through a reaction-diffusion based approach, focusing on the coupled interactions of four key variables - solid mass density, hydrolysed mass density, acetogenic mass density and methanogenic mass density, that themselves are stochastically affected by fluctuations, coupled with diffusive relaxation of the individual densities, in ambient surroundings. Our results indicate that close to the linearly stable limit, the large time steady state properties, arising out of a series of complex coupled interactions between the stochastically driven variables, are scarcely affected by the biochemical growth-decay statistics. Our results clearly show that an equilibrium landfill structure is primarily determined by the solid and hydrolysed mass densities only rendering the other variables as statistically \enquote{irrelevant} in this (large time) asymptotic limit. The other major implication of incorporation of stochasticity in the landfill evolution dynamics is in the hugely reduced production times of the plants that are now approximately 20-30 years instead of the previous deterministic model predictions of 50 years and above. The predictions from this stochastic model are in conformity with available experimental observations.  
\end{abstract}


\end{frontmatter}


\newpage
\section{Introduction}
\label{intro}

\noindent
Municipal waste management (MSW) has traditionally been a supply chain based facility primarily focused on ascertaining the most cost effective way of disposing household waste, including bio-waste. The concept of a modern landfill though stems from the idea of not only cost optimising bulk bio-waste disposal, but also to recycle the bio-disposables to convert chemical energy to industrially usable electric energy. From a supply chain perspective, this constitutes a feedback architecture where the disposable waste produces usable energy that is then fed back to the system itself for self-sustenance of the energy production process while simultaneously trafficking the extra energy generated for industrial usage \cite{Scheutz2009}. Such operational management of power production from disposable bio-waste fundamentally relies on the engineering novelty that could ensure maximum energy production at minimum bio-filler consumption while also maximising the profit generated by appropriate disbursement of the energy through the associated supply chain network \cite{Bariaz2012}. The success of such an \enquote{alternative energy} based industry then is inherently determined by the accuracy at which the following two factors can be probabilistically evaluated - the start time of the production process and the end time line up to which bulk production can be ensured from a plant.

Collection rates of the output landfill (methane) gas and associated collection efficiency are pivotal in quantifying the quality of a production plant and in future planning deliverables based on such production. Results at real methane production sites (methanogenic phase) have shown that the production rate and volume could drastically change depending on the nature and quality of clay covers, geosynthetic clay liners and geomembrane composite covers with the ${{\text{CH}}}_4$ \cite{Spokas2006} emission rates varying from 2.2 to 10,000 mg/${m}^{2}$/d. Aside of the core landfill engineering, alternative (methane) production methods in the form of microbial oxidation have been proposed as a cost efficient measure \cite{Humer2008}. Numerical models, focusing on the methane production rate with respect to the height dependence of landfill sites have supported such observations \cite{Gottinger1986, Li2002, Alidi1996, Perera2002} with additional information such as 99\% of the methane gas flow at the bottom being oxidized across the 0.8 m soil compost column with bulk oxidation of methane occurring within the top 0.2 m. The aspects of municipal solid waste management  \cite{Zhang2010,Pires2011} have been areas of recent research interest, especially focusing on landfilling impacts \cite{Elfadel1997} and systems analytics \cite{Pires2011,Chang1996} based perspectives. Such statistical studies have made extensive use of linear programmming algorithms \cite{Chang1996,Lia2006} in analyzing multi-phase mixing of leachates. In a recent work \cite{Gioannis2008}, landfill gas generation data from residual municipal solid waste (RMSW) have been utilized to estimate the anaerobic gas generation rate constants. Without having been explicitly mentioned in this article \cite{Gioannis2008}, the numbers obtained (0.0347-0.0803 ${\text{y}}^{-1}$) seem clearly to indicate the importance of incorporation of stochasticity in the landfill gas related mathematical models, albeit the model applied specifically to aerobically stabilised MSW. Order of magnitude estimates made in the context of the United Kingdom landfill data also agree with such numbers \cite{Allen1997}. The methane production time lines of real plants as indicated by numbers in this article estimate time periods between 12.5-33 years \cite{Gioannis2008} that are obtained by inverting these rate constants \footnote{Production time periods are roughly equal to the inverse of the gas generation rate constants.}. An accurate estimation of methane production time lines from landfill sites have evaded estimations from available deterministic theoretical models \cite{cvetkovic1994,elfadel1997a,elfadel1997b,elfadel2000,eastman1981,zacharof2004} in which these numbers come grossly overestimated by up to 150\% further confirming the need for improved theoretical models. 

In all these aforementioned studies, explicit incorporation of stochasticity in the otherwise deterministic dynamics could prove useful in analyzing the production process as a function of time. This is where stochastic mathematical modelling of the degradation rates of landfilled waste and consequent emergence of the hydro-carbon gases (e.g. methane) from the facility assumes importance. While traditional models \cite{cvetkovic1994,elfadel1997a,elfadel1997b,elfadel2000,eastman1981,zacharof2004,Young1989} have been successfully able to predict the correct deterministic core of the processes defining a landfill facility, very few of these have made any explicit allusion to stochastic modelling. While some attempts \cite{elfadel1997a,elfadel1997b,elfadel2000,zacharof2004} have been made to analyze the four phased biosolid$\to$hydrolysed leachate$\to$acetogenic compounds$\to$biogas (methane) production process, the incorporation of an explicit stochastic uncertainty could emphasise the role of phase heterogeneity in the mathematical model. The descriptions in these models clearly indicate the understanding for the need of such additional contributions in the model but related attempts were restricted only to the \enquote{mean field} probabilistic model.

The premise of this article is to bridge this information gap between realistic stochastic fluctuations of variables as seen in actual landfill sites and assumed deterministic approximation of the same in theoretical modelling as have been done. The target is to ensure that not only qualitative facts concerning the landfill dynamics are correctly accounted for but also accurate quantitative estimates of decay times of gas production facilities be estimated from the theoretical model. This is the objective of this article and would be studied using well established reaction-diffusion formalism as detailed below.

The article is organised as follows. After the derivation and description of the core stochastic model in the following section (Section 2), the temporal dynamics will be analysed in details (Section 3) where the focus will be on autocorrelation functions, the squared terms of which will clone the stochastic temporal dynamics. This then will be followed by a conclusion section (Section 4) where a summary of the main results will be drawn.§ 

\section{Materials and Methods}
\label{materials_section}

This is a paper on theoretical modelling of a stochastically forced dynamical process. As indicated already, this is a non-reductionist scheme geared toward optimised management of resources related to a waste management site, primarily focusing on the linear kinetics of the individual variables - solid mass density (${n_1}^{(s)}$), hydrolysed mass density (${n_2}^{(h)}$), acetogenic mass density (${n_3}^{(a)}$) and methanogenic mass density (${n_4}^{(m)}$) - together with their mutual coupled statistics that are often stochastically perturbed by the surrounding environment as well as through spatiotemporal parametric fluctuations. This will be a non-conventional approach that will address the role of complexity in the (hierarchical) biochemical pathways' network. Unlike its predecessors \cite{cvetkovic1994, elfadel1997a,elfadel1997b,elfadel2000,eastman1981} that have generally focused on data based phenomenological models, our continuum model will be structured around the well established reaction-diffusion scheme that has so often been successfully employed in addressing problems in biology \cite{murray,akc1}, fluid mechanics \cite{akc2} as well as in optimization of resources \cite{Alidi1996}. An intrinsic strength of our model is its ability to identify and analyse the origin of spatial heterogeneity and its time evolution with respect to interacting variables. As will be later seen, this is a key issue in determining the eventual decay times of the variables as has been shown elsewhere \cite{akc3}. A reaction-diffusion model with additive noise but with mutually coupled variables is equivalent to a multiplicative noise based time evolution that may lead to certain features of nonlinearity starting from a linear deterministic structure \cite{akc3}. One must also note that the choice of the nature of the noise distribution, that is whether it is \enquote{white} or \enquote{colored}, will quantitatively affect the dynamical process. At time scales close to the equilibrium time scale, though, this is not expected to affect the result by more than a few percentage at the most. The other tacit assumption, and hence a \enquote{weakness}, is the assumption of continuum dynamics. Once again, for all known practical purposes, this is not expected to have any major effect on the outcome.

\section{Theory}
\label{theory_section}
In line with the scheme enunciated in earlier works \cite{eastman1981,zacharof2004}, we introduce a generalised hydrolysing stoichiometric growth/decay rate as the solution of first order reaction-diffusion kinetics as follows

\begin{equation}
R(t) = At^\alpha \exp(-kt),
\label{rate_eqn}
\end{equation}

where $R(t)$ = reaction rate at time $t$ (kilogram/year) while $A$ and $k$ are respectively the amplitude and decay rates expressed in non-dimensional units. The constant $\alpha$ subjectively characterises the specific landfill concerned ($\alpha=1$ represents an earlier model \cite{zacharof2004} while other limiting values of $\alpha$ correspond to other models detailed before \cite{elfadel1997b,elfadel2000,eastman1981}). 

\subsection{Reaction-Diffusion Model}
\label{rd_section}

The first-order reaction-diffusion kinetics detailed in this section can now be engrained in the core stochastic reaction-diffusion model \cite{akc3,akc4} defined in line with the Langevin formalism \cite{akc5,Risken} as follows:

\begin{subequations}
\begin{equation}
\frac{\partial {n_1}^{(s)}}{\partial t} = \nu_1 \frac{\partial^2 {n_1}^{(s)}}{\partial {{\bf x}}^2} - k_h {n_1}^{(s)} + \eta_1({\bf x},t)
\label{n1_eqn}
\end{equation}

\begin{equation}
\frac{\partial {n_2}^{(h)}}{\partial t} = \nu_2 \frac{\partial^2 {n_2}^{(h)}}{\partial {{\bf x}}^2} + k_h {n_1}^{(s)} - A_a t^\alpha \exp(-k_a t) + \eta_2({\bf x},t)
\label{n2_eqn}
\end{equation}

\begin{equation}
\frac{\partial {n_3}^{(a)}}{\partial t} = \nu_3 \frac{\partial^2 {n_3}^{(a)}}{\partial {{\bf x}}^2} + {k_h}' {n_4}^{(m)} + A_a t^\alpha \exp(-k_a t) -  A_m t^\alpha \exp(-k_m t) + \eta_3({\bf x},t)
\label{n3_eqn}
\end{equation}

\begin{equation}
\frac{\partial {n_4}^{(m)}}{\partial t} = \nu_4 \frac{\partial^2 {n_4}^{(m)}}{\partial {{\bf x}}^2} - {k_h}' {n_4}^{(m)} +  A_m t^\alpha \exp(-k_m t)+ n_{H_2S}(t) + \eta_4({\bf x},t).
\label{n4_eqn}
\end{equation}
\end{subequations}

\noindent
In the above description, $k_h$ represents the degradation rate of the solid phase while ${k'}_h$ represents the same for the biogas (methane) phase. The A's are the biochemical growth (decay) rate amplitudes and the k's represent the corresponding decay time scales. As mentioned earlier, the diffusion terms $\nabla^2 n_i$ reflect the force of homogeneity working against the evident heterogeneous phase mixture that is present in a landfill. Although $H_2S$ is an unavoidable end product, since it comes with low relative percentage contribution ($\sim$ 0.05\%), we would drop this term for subsequent calculations at this level of modelling.

Here the respective diffusion terms ($\frac{\partial^2 {n_i}}{\partial {{\bf x}}^2}$) ($i$ = 1, 2, 3 and 4; the superscripts have been neglected to simplify notation, a convention that we will follow throughout this article henceforth) represent the heterogeneous relaxation of each variable ($n_1({\bf x},t) \neq n_2({\bf x},t) \neq n_3({\bf x},t) \neq n_4({\bf x},t)$) defined through diffusion constants $\nu_i$'s while the $\eta_i$'s characterise uncorrelated Gaussian white noises (an assumption) in three spatial dimensions corresponding to each variable $n_i$, as follows:

\begin{equation}
<\eta_i({\bf x},t) \eta_j({\bf x'},t')> = 2D_i \delta^3({\bf x-x'}) \delta(t-t') \delta_{ij}.
\label{noise_eqn}
\end{equation} 

The above continuum coupled model represented by equations (\ref{n1_eqn},\ref{n2_eqn},\ref{n3_eqn},\ref{n4_eqn}) abide the ensemble averaged mass-balance relation: $\frac{\partial}{\partial t}<[{n_1}^{(s)} + {n_2}^{(h)} + {n_3}^{(a)} + {n_4}^{(m)}>=0$, the curly bracket \enquote{$<>$} representing the ensemble average over all noise realisations. 

It must be noted that at this minimalist level, we are neglecting effects from noise cross-correlations (e.g. $<\eta_1 \eta_2> = 0$), an assumption which implies that the origin of stochasticity in solid mass density will not incite an identical stochastic response in the hydrolysed, acetogenic or methanogenic mass densities, a reasonable assumption at this stage.

In what follows, we will calculate the ensemble averaged root-mean-square values of the respective autocorrelation functions like $\sqrt{<n_i({\bf x},t) n_i({\bf x},t+\tau)>}$ in the large time ($t\to \infty$) steady state equilibrium limit. These autocorrelation functions represent the experimentally measured stochastic equivalents of their deterministic counterparts (as in \cite{zacharof2004}, only for $\nu_i=0$ though).

\section{Results}
\label{results_section}

As detailed in the previous section \ref{rd_section}), we will start with equations (\ref{n1_eqn},\ref{n2_eqn},\ref{n3_eqn},\ref{n4_eqn}) and then analyse these equations in the $k-\omega$ Fourier space. 

Starting with the the $n_1$-equation, we get

\begin{eqnarray}
\frac{\partial {n_1}}{\partial t} &=& \nu_1 \frac{\partial^2 {n_1}}{\partial {{\bf x}}^2} - k_h {n_1} + \eta_1({\bf x},t) \nonumber \\
-i\omega \tilde {n_1}({\bf k},\omega) &=& -\nu k^2 \tilde{n_1} - k_h \tilde{n_1} + \tilde{\eta_1} \nonumber \\
\tilde{n_1} &=& \frac{\tilde{\eta_1}({\bf k},\omega)}{-i\omega+\nu_1k^2 +k_h},
\end{eqnarray} 

followed by 

\begin{equation}
C_1(\tau) = <n_1({\bf x},t)* n_1({\bf x},t+\tau)> = \int d^3k \int d{\omega}\: e^{-i\omega \tau}\;<\tilde{n_1}({\bf k},\omega)\tilde{n_1}(-{\bf k},-\omega)>.
\end{equation}

In the experimentally viable large time equilibrium limit ($t\to \infty$), this leads to the following solution for the $n_1$-autocorrelation function:

\begin{subequations}
\begin{eqnarray}
<n_1({\bf x},t)* n_1({\bf x},t+\tau)> &=&  4\pi^2 D_1 \int_0^\infty dk\:\frac{k^2 e^{-(\nu_1 k^2 + k_h)\tau}}{\nu_1 k^2 + k_h} \label{n1_exact}\\
&\approx&  \frac{4\pi^2 D_1}{\nu_1}[\frac{\sqrt{\pi}}{2\sqrt{\nu_1 \tau}}\exp(-k_h \tau) + \frac{\pi k_h}{2}\frac{e^{-(1-\nu_1)k_h \tau}}{\sqrt{\nu_1 k_h}}],\:\:\text{for}\:\:(\nu_1<1).
\label{n1_autocorr}
\end{eqnarray}
\end{subequations}

\noindent
The above form given in equation (\ref{n1_autocorr}) is based on a complex integration within aforementioned limits of the wave vector k. For a more accurate expression valid for all k-limits, we will use the form given in equation (\ref{n1_exact}). This is the formula used in the plots shown later. 

Starting from equation (\ref{n2_eqn}), we get

\begin{equation}
-i\omega \tilde{n_2}({\bf k},\omega) = -\nu_2 k^2 \tilde{n_2} + k_h \tilde{n_2} -A_a h({\bf k},\omega) + \tilde{\eta_2}({\bf k},\omega),
\end{equation}

where $h({\bf k},\omega) = \frac{1}{\sqrt{2\pi k_a}}\Gamma(1+\alpha) e^{i(1+\alpha)\arctan(\frac{\omega}{k_a})}\:{(1+\frac{\omega^2}{{k_a}^2})}^{-\frac{1}{2}(1+\alpha)}\:(-{k_a}^{2\alpha})[-{k_m}^\alpha +{(-1)}^{1+\alpha}{k_a}^\alpha]$.

Without much loss of generality we may use the value $\alpha=1$ as in \cite{zacharof2004} to get $h({\bf k},\omega)=i\sqrt{2\pi}\frac{\delta(\omega+ik_a)}{\omega+ik_a}$, where \enquote{$\delta$} alludes to the celebrated Dirac-Delta function as is widely known in the literature \cite{akc1,akc2}.

The above prescription leads to

\begin{equation}
\tilde{n_2}({\bf k},\omega) = \frac{k_h \tilde{\eta_1}({\bf k},\omega)}{(-i\omega+\nu_2 k^2)(-i\omega +\nu_1 k^2 +k_h)} - \frac{i\sqrt{2\pi}A_a \delta(\omega+ik_1)}{(\omega+ik_a)(-i\omega+\nu_2 k^2)} + \frac{\tilde{\eta_2}({\bf k},\omega)}{-i\omega+\nu_2 k^2}
\end{equation}

that in turn gives

\begin{equation}
<\tilde{n_2}({\bf k},\omega)*\tilde{n_2}(-{\bf k},-\omega)> = \frac{k_h <\tilde{\eta_1}({\bf k},\omega)*\tilde{\eta_1}(-{\bf k},-\omega)}{(\omega^2+{\nu_2}^2 k^4)(\omega^2 +{(\nu_1 k^2 +k_h)}^2)} - \frac{2\pi {A_a}^2 \delta(\omega+ik_1)*\delta(-\omega+ik_a)}{(\omega^2+{k_a}^2)(\omega^2+{\nu_2}^2 k^4)} + \frac{\tilde{\eta_2}({\bf k},\omega)*\tilde{\eta_2}(-{\bf k},-\omega)}{\omega^2+{\nu_2}^2 k^4}.
\label{n2k_corr}
\end{equation}

As previously, for $\nu_1>\nu_2$, the above equations (\ref{n2k_corr}) may be complex integrated around all 5 poles in any of the halves of the respective Argand diagram to obtain
\begin{eqnarray}
<n_2({\bf x},t) *n_2({\bf x},t+\tau)> &=& \int d^3k \int d{\omega}\:e^{-i\omega \tau}\:
<\tilde{n_2}({\bf k},\omega)*\tilde{n_2}(-{\bf k},-\omega)> \nonumber \\
&=& \int_0^\infty dk\:\{\frac{2\pi D_1 {k_h}^2}{\nu_2(\nu_1 k^2 +k_h)}[\frac{1}{(\nu_1+\nu_2) k^2 k_h}(e^{-(\nu_1 k^2 +k_h)\tau}\nonumber \\
&+& e^{-\nu_2 k^2 \tau})+\frac{1}{(\nu_1-\nu_2)k^2 k_h}(e^{-\nu_2 k^2 \tau}-e^{-(\nu_1 k^2 + k_h)\tau})] + \frac{4\pi D_2}{\nu_2} e^{-\nu_2 k^2 \tau}\}
\label{n2_autocorr}
\end{eqnarray}

In line with derivations for the autocorrelation functions corresponding to variables $n_1$ and $n_2$, it should be noted that for the other two variables, our model defines the variable $n_4$ as the independent one while $n_3$ depends on $n_4$. Starting from equations (\ref{n3_eqn}) and (\ref{n4_eqn}) and following similar algebra as before, we can now evaluate the corresponding autocorrelation functions as follows:

\begin{eqnarray}
<n_4({\bf x},t) *n_4({\bf x},t+\tau)> &=& \int d^3k \int d{\omega}\:e^{-i\omega \tau}\:
<\tilde{n_4}({\bf k},\omega)*\tilde{n_4}(-{\bf k},-\omega)> \nonumber \\
&=& 4 \pi^2 D_4 \int_0^\infty dk\:\frac{k^2 e^{-(\nu_4 k^2 + {k_h}')\tau}}{\nu_4 k^2 + {k_h}'} \nonumber \\
&\approx&  \frac{4\pi^2 D_4}{\nu_4}[\frac{\sqrt{\pi}}{2\sqrt{\nu_4 \tau}}\exp(-{k_h}' \tau) + \frac{\pi {k_h}'}{2}\frac{e^{-(1-\nu_4){k_h}' \tau}}{\sqrt{\nu_4 {k_h}'}}],\:\:\text{for}\:\:(\nu_4<1).
\label{n4_autocorr}
\end{eqnarray}

In the limit ${k_h}' \to 0$, the above correlation function takes the limiting value ${<n_4({\bf x},t) *n_4({\bf x},t+\tau)>|}_{{k_h}' \to 0} = \frac{2 D_4 {\pi}^{3/2}}{{\nu_4}^{3/2} \sqrt{\tau}}$.

Using the information from equation (\ref{n4_autocorr}) above, the autocorrelation function defining the acetogenic decay dynamics in the limit ${k_h}' \to 0$ can be obtained as follows

\begin{eqnarray}
<n_3({\bf x},t) *n_3({\bf x},t+\tau)> &=& \int d^3k \int d{\omega}\:e^{-i\omega \tau}\:
<\tilde{n_3}({\bf k},\omega)*\tilde{n_3}(-{\bf k},-\omega)> \nonumber \\
&=& 4 \pi^2 [{{k_h}'}^2 D_4 \int_0^\infty dk\:\frac{k^2 e^{-((\nu_4-\nu_3)k^2+{k_h}')\tau}}{[(\nu_4+\nu_3)k^2+{k_h}'][(\nu_4-\nu_3)k^2+{k_h}']}+\frac{D_3}{\nu_3}(k_{\text{max}}-k_{\text{min}})],
\label{n3_autocorr}
\end{eqnarray}
where $k_{\text{max}}$ and $k_{\text{min}}$ are respectively defined as the inverse of the smallest and largest length scales in the problem. In the context of the landfill model, $k_{\text{min}}$ is the inverse of the landfill diameter while $k_{\text{max}}$ is the inverse of the landfill height. So the difference is a small finite number.

\section{Discussions}
\label{discussions_section}

The formulas presented above depict the complete dynamical behavior of the spatially heterogeneous landfill mechanism that, in turn, is governed by four sub-processes. In order to present the results quantitatively, in the above and all future formulations, we will consider same noise strengths, that is $D_1=D_2=D_3=D_4=D_0$ without any loss of generality.  
Using $\nu_1=1.0,\:\nu_2=0.8, \:k_h=1.0,\:D_1=0.1,\:D_2=0.05,\:{k_h}' \to 0$ as the parameter values, the solutions of eqn(\ref{n1_autocorr}) and eqn(\ref{n2_autocorr}) when plotted gives the following time decay profiles. 
It is to be noted that an inherent part of this conclusion relies on the fact that $\nu_1 \neq \nu_2$; in other words, on a heterogeneous spread through diffusional relaxation. 

The appended figures compare the root-mean-squared profiles of all four variables after averaging over all stochastic realisations. More specifically, we plot ${n_i}^{\text{rms}}(\tau) = \sqrt{<n_i({\bf x},t)*n_i({\bf x},t+\tau)>}$ (i=1,2,3,4) against the time difference $\tau$. As can be seen, this are {\it Gaussian stationary processes} \cite{akc4} implying that in the large time equilibrium limit ($t \to \infty$), the respective autocorrelation functions depend only on the time difference between two specific points of measurement and not on the time lines themselves.

\begin{figure}[htbp]
\centering
\includegraphics[scale=0.6]{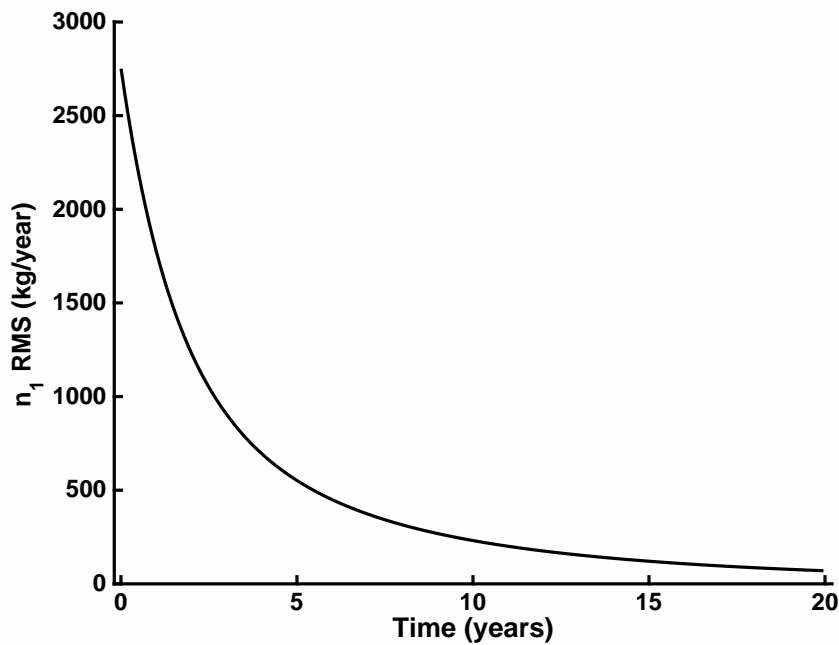}
\caption{The solid line represents the decay profile of the root mean square solid mass density autocorrelation function (represented by eqn(\ref{n1_autocorr})) with time.}
\label{n1plot}
\end{figure}

\begin{figure}[htbp]
\centering
\includegraphics[scale=0.6]{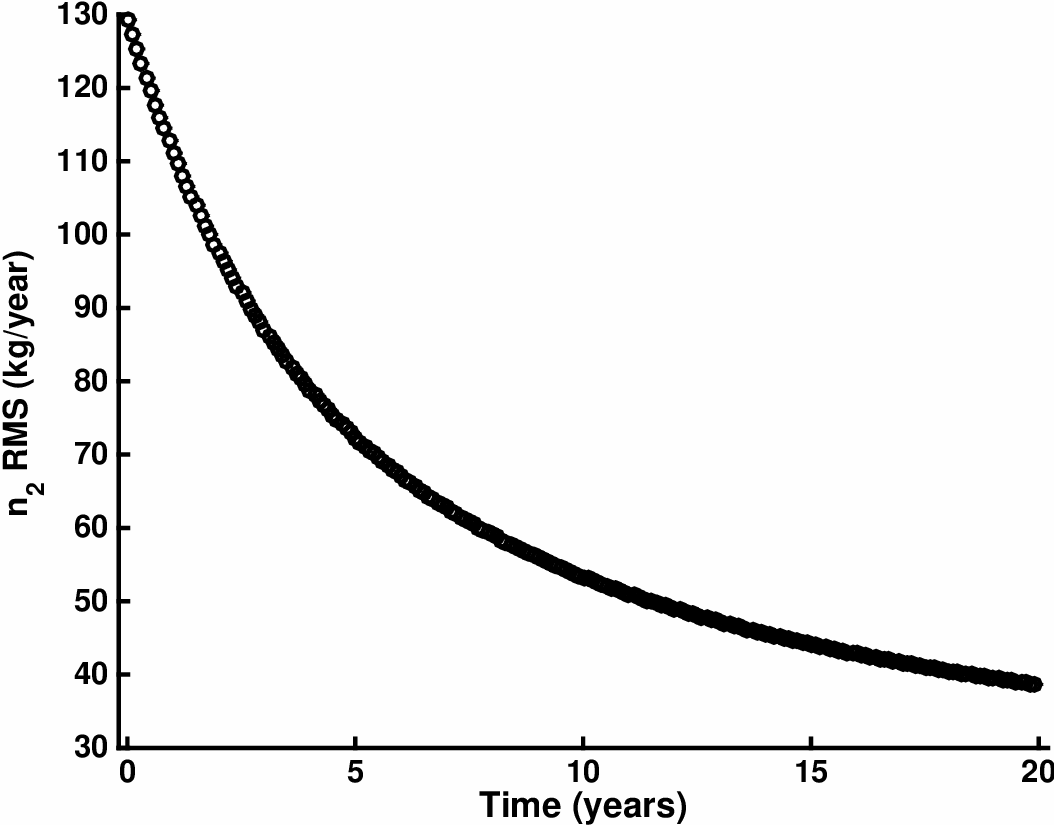}
\caption{The solid line represents the decay profile of the root mean square hydrolysed density autocorrelation function (represented by eqn(\ref{n2_autocorr})) with time.}
\label{n2plot}
\end{figure}

\begin{figure}[htbp]
\centering
\includegraphics[scale=0.6]{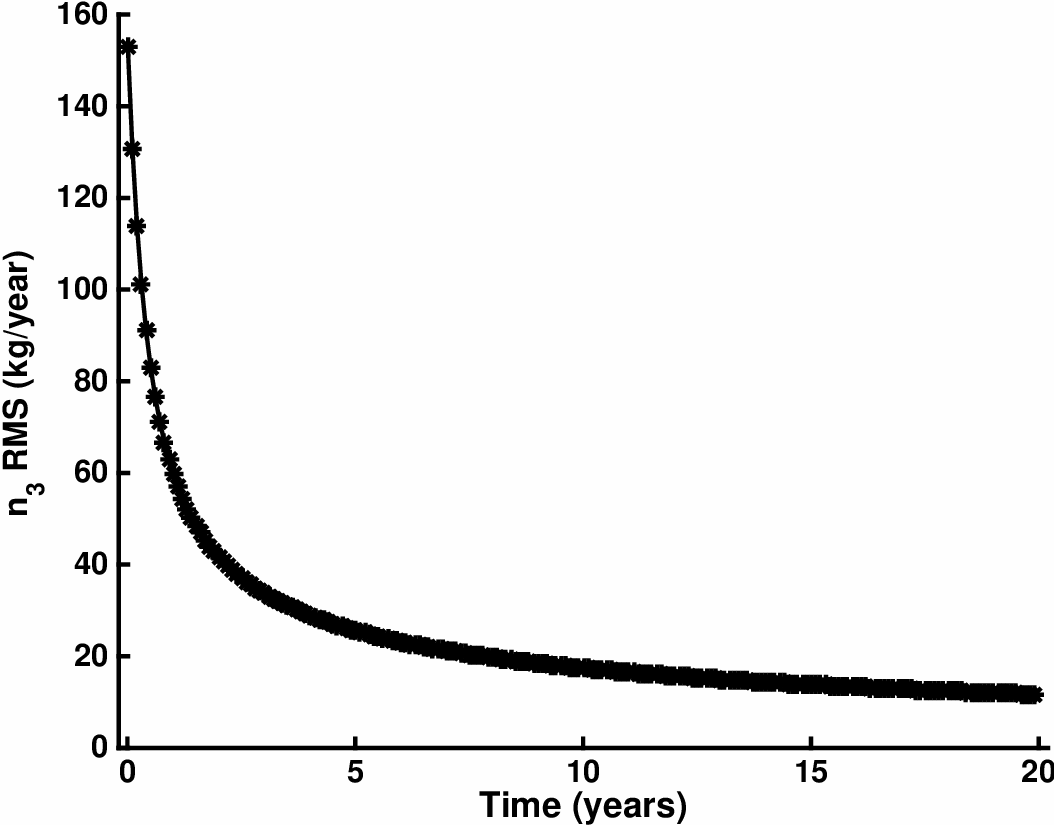}
\caption{The solid line represents the decay profile of the root mean square solid acetogenic density autocorrelation function (represented by eqn(\ref{n3_autocorr})) with time.}
\label{n3plot}
\end{figure}

\begin{figure}[htbp]
\centering
\includegraphics[scale=0.6]{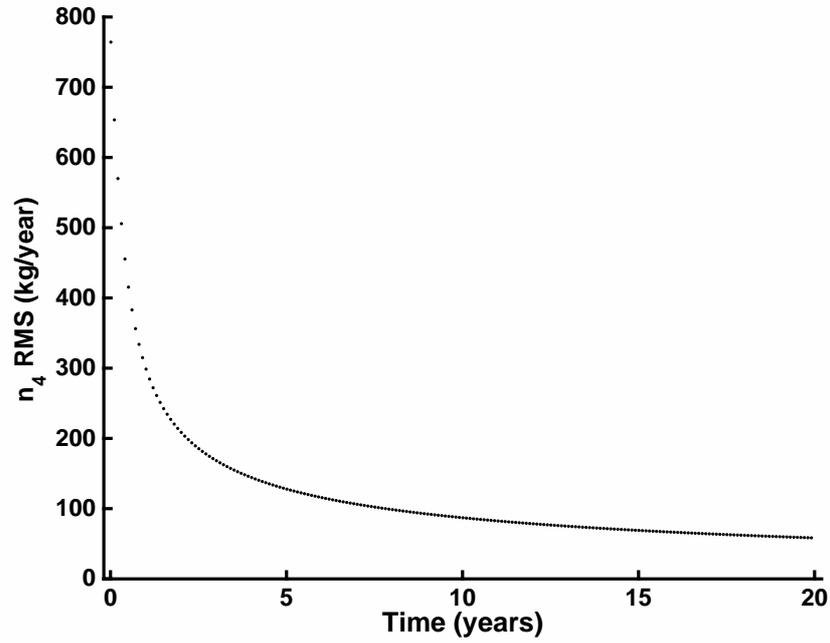}
\caption{The solid line represents the decay profile of the root mean square solid methanogenic density autocorrelation function (represented by eqn(\ref{n4_autocorr})) with time.}
\label{n4plot}
\end{figure}

\begin{figure}[htbp]
\centering
\includegraphics[scale=0.8]{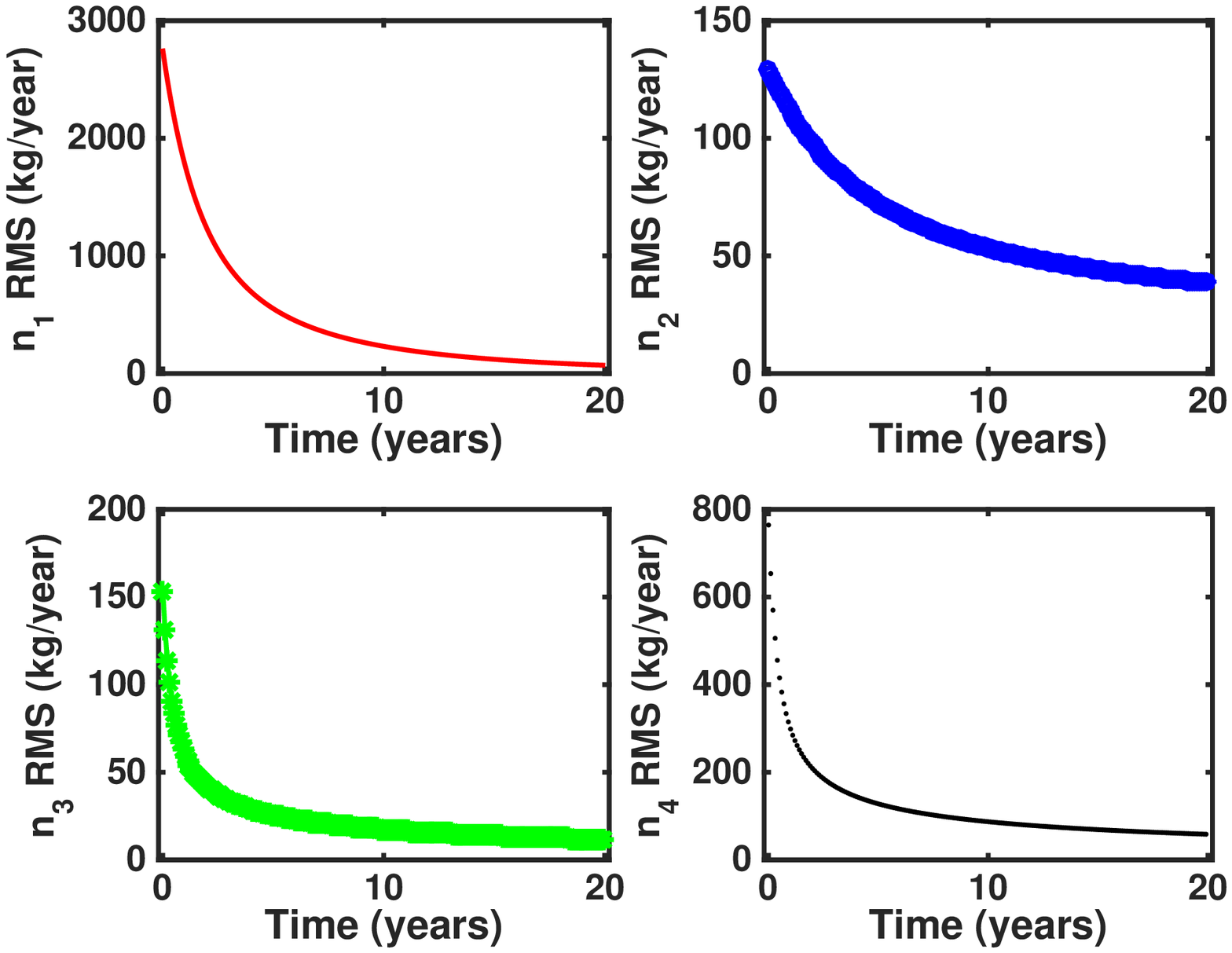}
\caption{Comparison of time variations of the mass densities of all four variables.}
\label{nallplot}
\end{figure}

Fig \ref{n1plot} and Fig \ref{n2plot} show the time decay of solid and hydrolysed waste that are used to generate the acetogenic and methanogenic phases shown in Fig \ref{n3plot} and Fig \ref{n4plot}. As is to be expected, the solid waste decay rate is steeper than the hydrolysed phase indicating that follow-up (methane) production necessitates a slow build-up leading to the target deliverables. 

The comparison of all four plots shown in Fig \ref{nallplot} encapsulates the summary of this theoretical model. While the decay rates of the individual phases vary, the final decay lines for all phases converge to the golden number of 20 years for all estimated phases. This number is subject to the parameter values used (indicated above). We have tested for other realistic parameter values availed from other publications \cite{Gioannis2008} (although this result is for aerobically stabilised MSW only, and hence is an order of magnitude comparison only) to confirm that the decay time line \footnote{The Gioannis, et al result is used for order of magnitude comparison only, since ours is an anaerobic model as opposed to the aerobically stabilised model of Gioannis, et al.} always conforms to the time window of 15-30 years which is in line with this reference. An understanding of this result can be drawn from the fact that our model simultaneously combines stochastic forcing with diffusive relaxation of the different densities, together which contribute to this experimentally agreeable timeline. This result is a huge improvement on existing deterministic models \cite{zacharof2004,Young1989}.

\section{Conclusions}

The implication of this theoretical analysis goes beyond the estimation of accurate gas production decay times and favourable comparisons with experimental data. The major benefit of such a study will be in the development of a robust business model in which quantitative dependence of such decay rates with varying system parameters will be part of the analytical description henceforth. As is not so very difficult to conceive, landfill site structures and engineering depend on the ambience and country specific facilities that may imply wide variation in parameter values. The results presented here incorporate all such provisions, including fluctuating parameter values. A pragmatic underpinning with regard to landfill engineering will be the precise quantitative control of parameters and clear ideas about the right parametric regime that will ensure gas production at a specified rate. As these rates may vary between different sites, as also with the country concerned, such numerical control would ensure easier and more direct improvement of existing landfill engineering frameworks. The work will also help municipalities and city councils to make right decision for landfill gas mining and implementing a sustainable landfill gas extraction as well energy recovery project. We believe that the relevance of this work can be best availed in association with national and supranational waste management policy makers. This may be particularly relevant in the context of a circular economy package, the focus of the European policy agenda (for instance), that is always more oriented towards landfill diversion and the promotion of other disposal options.

From the technical perspective, this analysis informs us that our stochastic linearly stable model is primarily affected by the time dependent hydrolysable decay rate ($R(t)$) at finite time scales $t<10$ years). For larger times, the steady state statistics remains unchanged with respect to changes in $R(t)$. As indicated above, the steady methanogenic gas production rate is primarily determined by the production rate of hydrolysed mass starting from solid mass with the acetogenic density contributing the least in the process. 
Unlike the previous models \cite{zacharof2004, Young1989}, our model satisfies the mass-balance relation at an ensemble averaged level and not for all deterministic realisations. This ensures dynamical equilibrium for all finite times: $<\frac{\partial}{\partial t}(n_1+n2+n_3+n_4)>=0$. A conclusion that we draw is that of higher decay rate for the solid mass density compared to the other three decay processes that is comparable to existing industrial results \cite{industry}.
In arriving at the plots, although specified fixed values of the noise strengths were used, but the linear stability ensured that the qualitative deductions obtained from the respective autocorrelation functions remain unaffected by the noise strengths.
Overall, it is predicted that the decay times of any of the four phases is shorter than the corresponding homogeneous deterministic case studied in previous models \cite{elfadel1997b,elfadel2000,eastman1981,zacharof2004}. All these above facts are directly related to numbers that a landfill engineer may make use of in designing the best landfill facility subjective to the given conditions. We should like to add that the model can be further extended to address specific structural remits, related to the legal paraphernalia of waste disposal in a country specific manner, and adjusting landfill plant lifelines in accordance with allowed range of plant structure parameters. The diffusion rates, which are key components in defining the landfill lifeline, are subjective of the size of the landfill; such legislations are often guided by European rules. Appropriate adjustment of the boundary conditions of our model could straight away address such components.

Another non-trivial aspect of this analysis is that the introduction of an explicit uncertainty in the dynamical model confirms the fact that heterogeneities ($\nu_1 \neq \nu_2 \neq 0$ and $\nu_3 \neq \nu_4 \neq 0$) play a vital role in the decay rate statistics. Our results predict that in absence of heterogeneity, mass distribution will take a far longer time. This was not so very obvious in most of the previous models. 

While retaining the core deterministic dynamics, that happens much to be the same in these models \cite{elfadel1997a,elfadel1997b,elfadel2000,zacharof2004}, our novelty lies in the introduction of two key factors that remained dormant in all these models. Firstly, explicit noise (stochastic) terms accompany each of our four dynamical equations representing each phase (biogas, hydrolysed leachate, acetogenic phase, methanogenous phase). Secondly, in order to incorporate the natural tendency of any physical system to neutralise the presence of any heterogeneity, we have incorporated diffusion terms in each of the phases that lead to a more generalised multiphase approach where the stochastically forced phases can mix with each other through diffusion.
n for the stochastic input; rather it is self-consistently derived from the original dynamics represents in equation(2a,2b,2c,2d). 

A reassuring quantitative confirmation of the analysis presented here comes from a comparison with real landfill descriptions as presented independently by Gioannis, {\it et al} \cite{Gioannis2008}. Inverting the decay constants estimated in this analysis, we reassuringly arrive at the production time line limitations as being between 12.5-33 years that is perfectly in harmony with numbers presented in this article (e.g. Fig \ref{nallplot}). This quantitative ramification also indicates the need for extending the scope of the present linearly stable model into the more realistic nonlinear regime; in other words strategy evaluation in planning waste management \cite{Li2012, Wilson1977}. This is the next plan along with a multivariate analysis of the outputs from the stochastic model based on a Fokker-Planck structure.

\section{Acknowledgment}
The authors acknowledge support of British Council funding under UKIERI thematic partnership 2012-15 in collaboration with Aston University, UK and Jadavpur University, India.








\end{document}